\documentclass[a4paper,12pt]{article}%

\usepackage{amsmath}
\usepackage{amsfonts}

\usepackage{amssymb}

\usepackage{bm}
\usepackage{amsthm}
\usepackage{amsfonts,amsmath,amssymb}

\title{Photon flux and distance from the source:
consequences for quantum foundations and technologies}

\author{Andrei Khrennikov, B\"orje Nilsson\\
International Center for Mathematical Modeling \\
in Physics, Engineering, Economics, and Cognitive Science\\
Department of Mathematics, Linn\ae us University \\SE-351 95 V\"{a}xj\"{o}, Sweden\\
Sven Nordebo\\
Department of Physics and Electrical Engineering\\ Linn\ae us University, SE-351 95 V\"{a}xj\"{o}, Sweden\\
Igor Volovich\\
Steklov Mathematical Institute, Russian Academy of Sciences\\ Gubkin St. 8, 119991, Moscow, Russia}

\begin{document}

\maketitle

\begin{abstract}
The paper explores the fundamental
physical principles of quantum mechanics (in fact, quantum field theory)
which limit the bit rate for long distances. Propagation of photons in optical fibers
is modeled using methods of quantum electrodynamics.
We define photon "duration" as the standard deviation of the photon arrival time;
we find its asymptotics for long distances and then obtain the main result of the paper:
the linear dependence of photon duration on the distance. This effect puts the limit
to joint increasing of the photon flux and the distance from the source and it has
 important consequences
both for quantum information technologies and quantum foundations.
Once quantum communication develops into a real technology, it would be appealing to the engineers
to increase both the photon flux and the distance. 
And here our ``photon flux/distance effect'' has to be taken into account  (at least if successively emitted
photons are considered as independent).  This effect also has to be taken into
account in a loophole free test of Bell's type -- to close jointly the detection and locality loopholes.
\end{abstract}


\section{Introduction}

Last years were characterized by a tremendous development of quantum information, in both theory and experiment. Quantum information
technologies, especially quantum cryptography, approached the level of  market products. Transmission of quantum information
for long distances is one of the most important problems of theoretical and experimental research \cite{UTS}. This problem has also a
foundational dimension as playing a crucial role in performance of a loophole-free test for Bell's type \cite{B} inequalities, see
\cite{Aspect}--\cite{Kwiat}.
Such a test should finally close all possibilities to interpret quantum mechanics as emergent from a local realistic model
(although, see, e.g., \cite{DEM4}-\cite{Groessing} for discussions, cf., e.g., \cite{Raedt}).\footnote{The present situation both in quantum foundations and quantum technologies (especially quantum cryptography and
quantum random generators)  is highly unsatisfying from the scientific viewpoint. There have been performed Bell's type tests closing
the locality and (recently) detection loopholes, separately, see \cite{Weihs}, \cite{Zeilinger}, \cite{Kwiat} (in this paper
we discuss only experiments with photons). The tests of both types demonstrated statistically significant  deviations from
the predictions of the local realistic model proposed by Bell \cite{B}, see \cite{KHR_CONT} for a discussion on generality of this model.
Often the results of these tests are interpreted as sufficient to reject local realism completely.} It is clear that
without a test which is free from every loophole, the present foundational grounds of quantum mechanics can be questioned. And it is not only the foundations that can be
questioned, but even the most successful quantum technologies such as quantum cryptography and quantum random generators.

As was already mentioned, recently two world leading experimental groups working in quantum foundations, first in Vienna \cite{Zeilinger} and then in
Urbana-Champaign \cite{Kwiat}, performed the Bell-type tests closing the detection loophole. Both groups have approached sufficiently
high levels of detection efficiency (which includes efficiency of detectors and all optical losses in the pathways from the source to
the detectors) to close the detection loophole, i.e., to discard the fair sampling assumption \cite{Adenier}. \footnote{
For us it is important to point to
the basic role of losses in the optical pathways, because the modern detectors which have been used both in the Vienna-experiment and
the Urbana-Champaign experiment have the efficiency near 100\%.} However, the locality loophole has not yet been closed jointly with the detection loophole
and nowadays the leading experimental groups work to improve the experimental setups from \cite{Zeilinger} and \cite{Kwiat}
(by increasing the distance between
detectors) in order to solve this problem.
The main difficulty is the essential increase in losses of photons with distance. Hence, it will be very difficult (if at all possible)\footnote{
In \cite{KV1} two authors of this paper formulated (on the heuristic grounds) a kind of complementarity
principle for closing the detection and locality loopholes.}    to
approach sufficiently high (for proceeding without the fair sampling assumption) arm-efficiency for long distances.

In this paper we study  spatial and temporal dependencies of detection probabilities for
photons propagating in optical fibers.
We define photon "duration" as the standard deviation of the photon arrival time;
we find its asymptotics for long distances and then obtain the main result of the paper:
{\it the linear dependence of photon duration on the distance.} This effect puts the limit
to joint increasing of the photon flux and the distance from the source and it has
 important consequences
both for quantum information technologies and quantum foundations.
In quantum technologies it is appealing to increase both the photon flux and the distance.
And here our ``photon flux/distance effect'' has to be taken into account  (at least if successively emitted
photons or pairs of entangled photons are treated as independent).  This effect also has to be taken into
account in a loophole free test of Bell's type -- to close jointly the detection and locality loopholes.
 As the result of increasing of losses in optical pathways, to collect sufficient statistical data
in this test, experimenters have to increase
duration of runs (for the fixed pairs of the orientations of polarization beam splitters). However, runs' durations cannot be too large, since
for very long runs it would be impossible  to neglect the drift effect, see \cite{Zeilinger1} for details.
Thus the photon flux has to be sufficiently high.
To exclude correlations between successively emitted pairs of entangled photons, experimenters have to select the photon flux 
compatible with photon duration which (as we shall show) increases linearly with the distance.

Now we outline the structure of the paper and its main result in details.
As was already pointed out, the main result of the paper is the derivation of the spatial asymptotics of the ``photon duration'',
 the standard deviation of instances of detection
of photons, see (\ref{10.5}).

In general, investigations of quantum correlations at long distances imply
the use of optical fibers. Nowadays, quantum communication links can be
established over considerable distances using the art fiber
technology \cite{PhysRevLett.87.167903,Landry:07,UTS,HVL}. Indeed, up
to 100 km entanglement distributions have been created in optical fibers
\cite{HVL,ZhangTakesueNamLangrockXieBaekFejerYamamoto2008,MarcikicdeRiedmattenTittelZbindenLegreGisin2004,Takesue2006,HonjoTakesueKamadaNishidaTadanagaAsobeInoue2007}%
.

The confinements of the photons to a fiber has sertain drawbacks. Firstly, there is a
loss of photons dependent mainly on material absorption, through internal
resonances or impurities, and Rayleigh scattering from local fluctuations in
material density\cite{Miya+Ternuma+Hosaka+Migashita1979}. Secondly, the fiber
is dispersive making the photon velocity frequency dependent. The photon loss can be reduced
by a high degree of purification of the fiber
material in combination with an optimal selection of the frequency band, whereas
the dispersion can be reduced by dispersion shift
\cite{Agrawal2010}. Even with such measures, the dispersion effect must usually
be taken into account for the design and evaluation of experiments on quantum
correlations. However, the loss of photons due to the mechanisms mentioned
above is often not critical and will be neglected in what follows.

One way of describing the effect of dispersion in the fiber is to concentrate on
scattering in the photon arrival time for an ensemble of photons having the
same initial state. In this paper, the standard deviation of photon arrival
time is denoted the "photon duration" time. This implies that in vacuum, where
dispersion is absent, the photon duration time vanishes. As an example Zhang
\emph{et al} \cite{ZhangTakesueNamLangrockXieBaekFejerYamamoto2008} report
that the photon duration time increases from 4 ps to 25 ps in a 100 km long
fiber, despite that both dispersion shift and a narrow frequency band have
been employed.

Expressions for the photon duration time based on quantum field theory are the
goal of the current paper. Their asymptotic form, valid for large distances,
should be useful for the design and evaluation of quantum correlation
experiments in optical fibers. A photon initial state serves as a model
for the generation of photons that usually is performed by down-conversion
using a laser and a non-linear crystal; see \emph{e.g.}
\cite{Christ+Laiho+Eckstein+Lauckner+Mosley+Silberhorn2009} for calculation of
the spatial modes produced in a non-linear waveguide.

A quantum field model of the fiber was given by Khrennikov et al. first for a
hollow waveguide \cite{OLD,KhrennikovNilssonNordeboVolovich2012} and then for a
single core optical fiber \cite{Khrennikov+Nilsson+Nordebo+Volovich2012} with the aim of
calculation of quantum correlations in a single discrete mode. As a
continuation of \cite{Khrennikov+Nilsson+Nordebo+Volovich2012}, the
quantization of a single discrete mode of the single core optical fiber was
re-examined, giving it a solid foundation by constructing the Hamilton
function in terms of cylindrical vector wave functions
\cite{Khrennikov+Nilsson+Nordebo+Volovich2012b}. This recent paper
\cite{Khrennikov+Nilsson+Nordebo+Volovich2012b} is the foundation of the
current paper.

The paper is organized so that it strictly defines (in section 2)
the photon duration time relating it to the probability density of finding a
photon. This probability density is then constructed in section 3 using the
quantization of a discrete mode of the single core fiber
\cite{Khrennikov+Nilsson+Nordebo+Volovich2012b} and then specialized to the
HE$_{11}$ mode that is usually used in single core fiber communication. The
expressions for the photon duration time given in sections 2-3 are simplified
in section 4 for the large distance regime. In the final section, a discussion
of the results is given.

\section{The photon arrival time and its standard deviation}

Consider a straight circularly symmetric (of radius $a>0$) fiber with the core and exterior domain
with material properties given by parameters $\mu_1, \varepsilon_1$ and $\mu_2, \varepsilon_2,$ respectively.
A cylindrical coordinate system $(\rho,\varphi,z)$ with
the $z-$axis in the middle of the core is attached to the fiber and the
coordinates are chosen so that the photons orginate inside the fiber
near the axial coordinate $z=0$ and time $t=0.$ The photon detection takes place in the
core at fixed axial coordinate: $z$ for one photon; $z_{1}$ and $z_{2}$ for a
photon pair. The arrival time is denoted $t(z)$ for a single photon and
$t_{1}(z_{1},z_{2}),t_{2}(z_{1},z_{2})$ for a photon pair. This paper deals
primarily with the mean value $\overline{t}(z)$ and the standard deviation
$\sigma(z)$ for a single photon arrival time.

$\sigma(z)$ is a measure on the scattering of the photon arrival time. A
natural name for $\sigma(z)$ is therefore
\cite{ZhangTakesueNamLangrockXieBaekFejerYamamoto2008,Agrawal2010} the photon
duration time. This notation should not be confused with the detection time,
the time it takes to detect the photon, which is neglected in this study. From
experimental values $\left\{  t_{n}(z)\right\}  _{n=1}^{N}$ of the photon
arrival time, the photon duration time $\sigma(z)$ can be estimated with%

\begin{equation}
\sqrt{\frac{1}{N-1}\left[  \sum_{n=1}^{N}t_{n}^{2}(z)-\frac{1}%
{N}\left(  \sum_{n=1}^{N}t_{n}(z)\right)  ^{2}\right]  }. \label{sigma}%
\end{equation}

A quantum electrodynamic field description of the fiber gives $P(\rho,z,t)$,
via Glaubert's formulae \cite{MW}, so that the probability of detecting a
photon with electrical polarization $\bm{\nu}$ in a specific discrete mode at
time $t$ in the volume element d$V$ at $(\rho,\varphi,z)$ is proportional to
$P(\rho,z,t)$d$V.$ In this section, formulae for $\overline{t}(z)$ and
$\sigma(z)$ are derived in terms of $P(\rho,z,t).$

The probability density $p(z,t)$ with respect to $t$ for locating a photon
anywhere in the core of the fiber at the cross-section located at $z$ is
\begin{equation}
p(z,t)=\frac{P(z,t)}{P_{\bm{\nu}}\int_{0}^{\infty}P(z,t)\mathrm{d}t},\label{4}%
\end{equation}
where $P_{\bm{\nu}}$ is the probability of finding a photon in polarization
$\bm{\nu}$ at any time, and
\begin{equation}
P(z,t)=2\pi\int_{0}^{a}P(\rho,z,t)\rho\,\mathrm{{d}\rho.}\label{5}%
\end{equation}
$P_{\bm{\nu}}$, which depends on the experimental set-up, is assumed to be a
constant, independent of the axial distance $z$. This means that the
probability for locating a photon anywhere in the core at the cross-section at
$z$ during the time interval $(t_{1},t_{2})$ is
\begin{equation}
\int_{t_{1}}^{t_{2}}p(z,t)\mathrm{{d}t.}\label{6}%
\end{equation}
Note that the joint probability density with respect to $t^{\prime}$ and
$z^{\prime}$ is $p(z,t^{\prime})\delta(z-z^{\prime})$.

Let
\begin{equation}
\mathcal{E}\{f(t)\}(z)=\int_{0}^{\infty}f(t)p(z,t)\text{d}t\label{7}%
\end{equation}
denote the expectation value of $f(t)$ with respect to the probability density
$p(t,z)$. With this notation, we get that the expectation value $\overline
{t}(z)$ of the arrival time of the photon is given by%

\begin{equation}
\overline{t}(z)=\mathcal{E}\{t\}(z)\label{7.5}%
\end{equation}
and the corresponding standard deviation
\begin{equation}
\sigma(z)=\sqrt{\mathcal{E}\{t^{2}\}(z)-\left[  \overline{t}(z)\right]  ^{2}}.\label{8}%
\end{equation}

It is convenient to express the results in terms of the moments
\begin{equation}
\tau_{n}(z)=\int_{0}^{\infty}t^{n}P(z,t)\text{d}t\label{9}%
\end{equation}
with respect to the un-normalized probability density $P(t,z)$, rather than
with respect to $p(z,t)$, giving
\begin{equation}
\left\{
\begin{array}
[c]{l}%
\overline{t}(z)=\dfrac{\tau_{1}(z)}{P_{\bm{\nu}}\tau_{0}(z)}\\[2ex]%
\sigma(z)=\sqrt{\dfrac{\tau_{2}(z)}{P_{\bm{\nu}}\tau_{0}%
(z)}-\left(  \dfrac{\tau_{1}(z)}{P_{\bm{\nu}}\tau_{0}(z)}\right)  ^{2}}%
\end{array}
\right.  .\label{10}%
\end{equation}
The explicit dependence on the polarization $\bm{\nu}$ $\ $and mode properties
for $\overline{t}(z)$ and $\sigma(z)$ are suppressed in what follows.

A major result of this paper is that%
\begin{equation}
\left\{
\begin{array}
[c]{l}%
\overline{t}(z)\sim Az\\
\sigma(z)\sim Bz
\end{array}
\right.  ,z\rightarrow\infty,\label{10.5}%
\end{equation}
where $A$ and $B$ are the explicit expressions independent of $z.~$This
means that asymptotically for large $z,$ the photon duration time $\sigma(z)$
increases linearly with $z.$ As a consequence, the measurement time like the
time required to determine probabilities also increases with $z.$ Conversely,
$\sigma(z)$ can be used to estimate the measurement time at a given distance
$z.$ An efficient method to determine $\sigma(z)\ $is provided by
(\ref{10.5}): first, the proportionality constant is determined for
sufficiently large $z=z_{0},$ then $\sigma(z)\sim Bz$ is calculated for
$z>z_{0}.$

\section{Quantization and the probability density}

In this section, the probability density for detecting a photon is
constructed. The basis for this is a quantization of a discrete mode in the single core
fiber \cite{Khrennikov+Nilsson+Nordebo+Volovich2012b}; all other discrete and
the non-discrete modes can be neglected for long distance interaction; see
\textit{e.g.}
\cite{Nordebo+Nilsson+Biro+Cinar+Gustafsson+Gustafsson+Karlsson+Sjoberg2013}.

Consider one mode with azimuthal index $m$ and let $|0\rangle$ be the
corresponding Fock vacuum with the creation operator $a_{mk}^{(+)};$ the
anihilation operator is $a_{mk}^{(-)}.$ With the initial state
\begin{equation}
|\Psi_{1}\rangle=\int_{\mathbb{R}}\mathrm{d}k\,g_{mk}a_{mk}^{(+)}|0\rangle,
\label{3}%
\end{equation}
the probability density for finding a photon in polarization $\nu$ is%
\begin{equation}
P_{\mathbf{\nu}m}(\rho,z,t)=\langle\Psi_{1}|\mathcal{E}_{\mathbf{\nu}m}%
^{(-)}(\rho,\varphi,z,t)\mathcal{E}_{\mathbf{\nu}m}^{(+)}(\rho,\varphi
,z,t)|\Psi_{1}\rangle, \label{1}%
\end{equation}
using Glaubert's formula \cite{MW}. The dependence on the polarization, which
was suppressed in the notation in section 2, is made explicit with a sub-index
$_{\nu}.$ According to \cite{Khrennikov+Nilsson+Nordebo+Volovich2012b}, the
electrical operator projected in direction $\mathbf{\nu}$ is%
\begin{equation}
\bm{\mathcal{E}}_{\nu m}^{(+)}(\rho,\varphi,z,t)=\int_{\mathbb{R}}%
\frac{\mathrm{d}k}{2\pi}\sqrt{\frac{\hbar\omega_{mk}}{2\epsilon_{0}}}%
a_{mk}^{(+)}\bm{\nu}\cdot\bm{\psi}_{mk}(\rho)\mathrm{e}^{\mathrm{i}%
(kz-\omega_{mk}t+m\varphi)}.\, \label{2}%
\end{equation}
Here, $\bm{\mathcal{E}}_{\nu m}^{(-)}(\rho,\varphi,z,t)$ is the Hermitian
conjugate of $\bm{\mathcal{E}}_{\nu m}^{(+)}(\rho,\varphi,z,t).$ $\omega_{mk}$
solves the dispersion relation
\begin{multline}
G_{m}(\omega,k)=\frac{a^{2}\kappa^{2}q^{2}}{k_{0}^{2}\mu_{1}\mu_{2}}%
\mathrm{J}_{m}^{2}(\kappa a)\mathrm{K}_{m}^{2}(qa)\left[  -\frac{m^{2}k^{2}%
}{k_{0}^{2}}\left(  \frac{1}{(qa)^{2}}+\frac{1}{(\kappa a)^{2}}\right)
^{2}\right. \label{eq:Gdet2}\\
\left.  +\left(  \frac{\mu_{1}}{\kappa a}\frac{\mathrm{J}_{m}^{\prime}(\kappa
a)}{\mathrm{J}_{m}(\kappa a)}+\frac{\mu_{2}}{qa}\frac{\mathrm{K}_{m}^{\prime
}(qa)}{\mathrm{K}_{m}(qa)}\right)  \left(  \frac{\epsilon_{1}}{\kappa a}%
\frac{\mathrm{J}_{m}^{\prime}(\kappa a)}{\mathrm{J}_{m}(\kappa a)}%
+\frac{\epsilon_{2}}{qa}\frac{\mathrm{K}_{m}^{\prime}(qa)}{\mathrm{K}_{m}%
(qa)}\right)  \right]  .
\end{multline}
where the $\omega$-dependence is introduced via $k_{0}=\omega\mu_{0}%
\epsilon_{0},$ $\kappa=\sqrt{k_{0}^{2}\mu_{1}\epsilon_{1}-k^{2}}$ and
$q=\sqrt{k^{2}-k_{0}^{2}\mu_{2}\epsilon_{2}}$. It is assumed that $\mu
_{1}\epsilon_{1}>\mu_{2}\epsilon_{2}$. Here, $\mu_{0}$ and $\epsilon_{0}$ are
the permeability and permittivity in vacuum, $\mu_{j}$ and $\epsilon_{j}$ the
relative permeability and permittivity in material $j$ with $j=1$ referring to
the core and $j=2$ to the exterior region, see Fig. 1. Explicit procedures for
calculating $\bm{\nu}\cdot\bm{\psi}_{mk}(\rho)$ are given in
\cite{Khrennikov+Nilsson+Nordebo+Volovich2012b}. From symmetry considerations
follows that $P_{\mathbf{\nu}m}(\rho,z,t)$ is independent of the azimuthal
angle $\varphi$ as shown in the notation. Only propagating modes are
considered and it is sufficient to consider non-negative $m$ so that
$\omega_{mk}=\omega_{m-k}$ is real and non-negative.

A regularization of the theory developed below for calculating $\overline
{t}(z)$ and $\sigma(z)$ will be required since $\omega_{mk}$ may vanish for
$k=0$. This is the case for the $\mathrm{HE}_{11}$ mode, with the asymptotic
behaviour $\omega_{mk}\sim|k|c_{0}(\epsilon_{2}\mu_{2})^{-1/2},k\rightarrow0$.
The regularization is done by replacing $k^{2}$ with $k^{2}+\varepsilon^{2}$
and let letting $\varepsilon$ tend to zero at the end of the analysis. Then
$\omega_{mk\varepsilon}$ is positive and satisfies $G_{m}(\omega
_{mk\varepsilon},\sqrt{k^{2}+\varepsilon^{2}})=0$ and $\omega_{mk}%
=\lim_{\varepsilon\rightarrow0}\omega_{mk\varepsilon}.$To assure that this
limit exists an initial state is selected so that $g_{mk}$ is vanishing
sufficiently fast when $k$ tends to zero.

The calculation of the moments $\tau_{n}(z)$ starts with standard procedures
with the commutation relations $a_{mk\varepsilon}^{(-)}a_{mk^{\prime
}\varepsilon}^{(+)}-a_{mk^{\prime}\varepsilon}^{(+)}a_{mk\varepsilon}%
^{(-)}=2\pi\delta(k-k^{\prime})\mathcal{I}$%
\cite{Khrennikov+Nilsson+Nordebo+Volovich2012b}$,$ with $\mathcal{I}$ being
the identity operator. Assuming a one photon state the result is
\begin{equation}
P_{\bm{\nu}m}(\rho,z,t)=\left\vert A_{\bm{\nu}m}(\rho,z,t)\right\vert
^{2},\label{11.5}%
\end{equation}
where
\begin{equation}
\left\{
\begin{array}
[c]{l}%
A_{\bm{\nu}m}(\rho,z,t)=\lim_{\varepsilon\rightarrow0}\int_{\mathbb{R}%
}\mathrm{d}k\,f_{\bm{\nu}mk\varepsilon}(\rho)\,\mathrm{e}^{\mathrm{i}%
(kz-\omega_{mk\varepsilon}t)}\\[1ex]%
f_{\bm{\nu}mk\varepsilon}(\rho)=g_{mk}\sqrt{\dfrac{\hbar\omega_{mk\varepsilon
}}{2\epsilon_{0}}}\bm{\nu}\cdot\bm{\psi}_{mk\varepsilon}(\rho).
\end{array}
\right.  \label{12}%
\end{equation}
The reality condition $f_{\bm{\nu}m-k\varepsilon}(\rho)=f_{\bm{\nu
}mk\varepsilon}^{\ast}(\rho)$ follows from the definitions of $g_{mk}$ and
$\bm{\psi}_{mk\varepsilon}(\rho)$.

\section{The photon duration time for large distances}

For a given initial state $|\Psi_{1}\rangle,$ defined by $g_{mk},$ the mean
value of the arrival time $\overline{t}(z)$ and the photon duration time
$\sigma(z)$ are given by (\ref{10}), (\ref{9}) and (\ref{5}) with
$P(\rho,z,t)=P_{\bm{\nu}m}(\rho,z,t)$ according to (\ref{11.5})-(\ref{12}).

It is assumed that
\begin{equation}
\omega_{mk^{\prime}\varepsilon}-\omega_{mk\varepsilon}=(k^{\prime
}-k)(k^{\prime}+k)F_{m\varepsilon}(k^{\prime},k),\varepsilon>0 \label{3.5}%
\end{equation}
with $F_{m\varepsilon}(k^{\prime},k)>0$, has zeros only at $k^{\prime}=\pm k$
and that these zeros are simple. This means that $\omega_{mk\varepsilon}$ is
increasing with $|k|$ with a minimum positive value at $k=0$. From
$\omega_{mk\varepsilon}=\omega_{m-k\varepsilon}$ and (\ref{3.5}) follows that
$F_{m\varepsilon}(k^{\prime},k)=F_{m\varepsilon}(k,k^{\prime})=F_{m\varepsilon
}(-k^{\prime},k)$. The assumption (\ref{3.5}) and $F_{m\varepsilon}(k^{\prime
},k)>0$ holds for the $\mathrm{HE}_{11}$ mode.

As a starting point for deriving asymptotic results for the $n^{\mathrm{th}}$
moment $\tau_{n}(z)$ when $z$ is large, (\ref{11.5}-\ref{12}) are inserted
into (\ref{9}) and the order of integration is changed to get
\begin{equation}
\tau_{n}(z)=\lim_{\varepsilon\rightarrow0}2\pi\int_{0}^{a}\rho\,\mathrm{d}%
\rho\int\!\!\!\int_{\mathbb{R}^{2}}\mathrm{d}k\,\mathrm{d}k^{\prime
}f_{\mathbf{\nu}mk\varepsilon}(\rho)f_{\bm{\nu}mk^{\prime}\varepsilon
}^{\ast}(\rho)\mathrm{e}^{\mathrm{i}(k-k^{\prime})z}I_{nm\varepsilon
}(k^{\prime},k).\label{13}%
\end{equation}
Here,
\begin{align}
I_{nm\varepsilon}(k^{\prime},k) &  =\int_{0}^{\infty}t^{n}\mathrm{{e}%
}^{\mathrm{{i}}(\omega_{mk^{\prime}\varepsilon}-\omega_{mk\varepsilon}%
)t}\mathrm{d}t\label{14}\\
&  =\frac{n!}{\left[  -\mathrm{{i}}(\omega_{mk^{\prime}\varepsilon}%
-\omega_{mk\varepsilon})\right]  ^{n+1}}+\pi(-\mathrm{{i}})^{n}\delta
^{(n)}(\omega_{mk^{\prime}\varepsilon}-\omega_{mk\varepsilon})\label{15}%
\end{align}
is a distribution meaning that the first term on the right hand side of
(\ref{15}) requires that the integration in $k$ and $k^{\prime}$ be
interpreted as Cauchy principle value integrals. Now, (\ref{3.5}) is
introduced into (\ref{15}). The integration in (\ref{13}) related to the
second term in (\ref{15}) can be performed using the definition of the
$\delta^{(n)}$ distribution whereas for the first term an asymptotic analysis
for large $z$, based on the standard Laplace transform
\begin{equation}
\int_{0}^{\infty}\ln t\,\mathrm{e}^{-st}\mathrm{d}t=-\dfrac{1}{s}\left(
\gamma+\ln s\right)  ,\label{16}%
\end{equation}
is appropriate, $\gamma=0.5772\ldots$ being Euler's constant. Let us define
\begin{equation}
\left\vert f_{\bm{\nu}mk\varepsilon}\right\vert ^{2}=2\pi\int_{0}^{a}%
\rho\,\mathrm{d}\rho\left\vert f_{\bm{\nu}mk\varepsilon}(\rho)\right\vert
^{2},\label{16.5}%
\end{equation}
which can be expressed in terms of Bessel functions using the results of
\cite{Khrennikov+Nilsson+Nordebo+Volovich2012b}. Then, the result is
\begin{equation}
\tau_{n}(z)=\tau_{n}^{\sim}z^{n}\left[  1+\mathrm{o}(1)\right]  ,z\rightarrow
\infty,\text{ }n=0,1,2,\label{17}%
\end{equation}
where the asymptotic constants
\begin{equation}
\left\{
\begin{array}
[c]{l}%
\tau_{0}^{\sim}=\lim_{\varepsilon\rightarrow0}\dfrac{\pi}{2}\int_{\mathbb{R}%
}\mathrm{d}k\dfrac{\left\vert f_{\bm{\nu}mk\varepsilon}\right\vert ^{2}%
}{|k|F_{m\varepsilon}(k,k)}\\[1.5ex]%
\tau_{1}^{\sim}=-\lim_{\varepsilon\rightarrow0}\dfrac{\pi}{4}\int_{\mathbb{R}%
}\mathrm{d}k\ln|2k|\dfrac{\mathrm{d}^{2}}{\mathrm{d}k^{2}}\dfrac{\left\vert
f_{\bm{\nu}mk\varepsilon}\right\vert ^{2}}{\left[  F_{m\varepsilon
}(k,k)\right]  ^{2}}\\[3ex]%
\tau_{2}^{\sim}=\lim_{\varepsilon\rightarrow0}\dfrac{\pi}{8}\int_{\mathbb{R}%
}\mathrm{d}k\dfrac{\left\vert f_{\bm{\nu}mk\varepsilon}\right\vert ^{2}%
}{|k|^{3}\left[  F_{m\varepsilon}(k,k)\right]  ^{3}}.
\end{array}
\right.  \label{18}%
\end{equation}
With (\ref{7.5}), (\ref{8}) and (\ref{17}) we finally have
\begin{equation}
\left\{
\begin{array}
[c]{l}%
\overline{t}(z)=\dfrac{1}{P_{\bm{\nu}}}\dfrac{\tau_{1}^{\sim}}{\tau_{0}^{\sim}}z+\mathrm{o}\left(
z\right)  ,z\rightarrow\infty\\[2ex]%
\sigma(z)=\sqrt{\dfrac{\tau_{2}^{\sim}}{P_{\bm{\nu}}\tau_{0}^{\sim}}-\left(  \dfrac
{\tau_{1}^{\sim}}{P_{\bm{\nu}}\tau_{0}^{\sim}}\right)  ^{2}}+\mathrm{o}\left(  z\right)
,z\rightarrow\infty.
\end{array}
\right.  \label{19}%
\end{equation}

Provided that the moments $\tau_{n}(z),$ $n=0,1,2$ exist, the variance
$\sigma^{2}(z)$ is non-negative that shows
that the asymptotic expression (\ref{19}) for $\sigma(z)$ is well defined. 

Asymptotic photon duration times for entangled states can also be derived straightforwardly, although the calculations are technically more involved. Only the results for the biphoton case are presented omitting all details in the derivation. An ensemble of identical initial states consisting of $2$ photons is considered. The two detectors are located at $z_1$ and $z_2$, respectively. For the duration time $\sigma_j(z_1,z_2)$ of photon $j$, defined as the standard deviation of the arrival time, we have
\begin{equation}\label{20}
\sigma_j(z_1,z_2)\sim B_j|z_j|,j=1,2~ {\rm when}|z_1|,|z_2|\rightarrow \infty.
\end{equation}
This means that the asymptotic one photon results (\ref{10.5}) is transferred to the two photon case provided that both observation points tends to $\infty$. Note that the interference terms found in the asymptotic formulae for the probability density, see (33) in \cite{Khrennikov+Nilsson+Nordebo+Volovich2012} for details, are not present in the lowest order asymptotic result (\ref{20}) for the duration time.

\section{Discussion}
\label{dis}

Our statistical theory is based on the assumption that there is an
ensemble of photons prepared in the same way. To test our predictions in
an experiment, there must be sufficiently long time interval between the emitted
photons, so that they can be considered independent.

In a real experiment one cannot wait an infinite time before emitting
the next photon. Still, the time interval between the successive emissions of
photons must be (in average) much larger than the
other time scales of the problem. One such time scale is the duration
time of the photon. If the time interval is not much larger than the
duration time of each photon then the two successively emitted photons (propagating in the same direction)
can no more be considered as independent systems with their own intrinsic properties;
they should be considered as an entangled photon pair. Thus, for measurements in an
optic fiber, the distance between the source and the detector comes into play if the
photons may interact; the longer the distance the more interaction there is
 for a given initial time interval. Hence,  two photons may be considered
as two independent systems on short distances, but 
we cannot assume that they are independent for sufficiently large distances 
(the photons may be entangled).

This inter-relation between the  photon flux
and the distance between the source and the detectors
is the main experimental consequence of our study.  It has important implications for quantum technologies and
quantum foundations.

Once quantum communication develops into a real technology, it would be appealing to the engineers
to increase both the photon flux and the distance. Our paper explores the fundamental
physical principles of quantum mechanics (in fact, quantum field theory) that limits the bit rate for long distances.

In coming experimental realizations of the loophole free Bell's test, see Introduction, the same problem arises.
On one hand, to close the locality loophole the distance has to be essentially increased comparing to the Vienna and Urbana-Champaign
experiments, \cite{Zeilinger}, \cite{Kwiat}. On the other hand, the photon flux has to be sufficiently high: otherwise
the duration of each run would become too large to neglect the drift effect, see \cite{Zeilinger1} for details.
Our paper says that one has to take into account {\it the photon flux/distance effect.}

In a real Bell's type experiment a laser beam is sent to a nonlinear crystal and  beams of photons go out from the crystal.
A beam has a  conical structure. One captures the photons in the beam with the end of the core of the optic fiber.
To do this, the beam must have sufficiently many photons in a small solid angle. Then,  it is natural to define
the brightness of the source as power/solid angle, while in quantum optics it may be convenient
to use the number of photons emitted each second (rather than power). It is also of importance to have nearly the same
energy or frequency of the photons in the beam. Of course, it is impossible that all photons  have precisely
the same frequency. Therefore, it is of interest to define the spectral brightness which is given as
$\rm{brightness} \times \rm{frequency}/\rm{width \; of \; frequency \; band}.$
Thus,  a source with a high spectral brightness at a particular angle and frequency has its photons concentrated to
this direction and this frequency. Both the photon flux and brightness are proportional to the number of photons emitted per
second (but the two notions are not identical). To get a closer connection to experiments testing
Bell inequalities,  in addition to the number of photons emitted each second, the spectral brightness is also useful.
Therefore, it may be convenient to speak not about the photon flux/distance effect, but {\it the
spectral brightness/distance effect.}

\medskip

{\bf Acknowledgment:}  The work was supported by the following funding agencies:
MPNS COST Action MP1006 (Fundamental Problems in Quantum Physics), Austrian Academy of Science (visiting professor fellowship of A. Khrennikov
at Vienna University and Atom Institute, 2013).


\end{document}